\RequirePackage{fix-cm}
\documentclass[twocolumn]{svjour3}          
\smartqed  
\usepackage{graphicx}
\journalname{Journal of Radioanalytical Nuclear Chemistry}
\begin{document}

\title{Investigation of production routes for the $^{161}$Ho Auger-electron emitting radiolanthanide, a candidate for therapy
}


\author{F. T\'ark\'anyi \and F. Ditr\'oi \and A. Hermanne \and S. Tak\'acs  \and A.V. Ignatyuk
}


\institute{F. Ditr\'oi F. T\'ark\'anyi \and S. Tak\'acs\at
              Institute of Nuclear Research of the Hungarian Academy of Sciences \\
              Tel.: +36-52-509251\\
              Fax: +36-52-416181\\
              \email{ditroi@atomki.hu}           
           \and
           A. Hermanne \at
              Cyclotron Laboratory, Vrije Universiteit Brussel (VUB), Brussels, Belgium
           \and
              A.V. Ignatyuk \at
            Institute of Physics and Power Engineering (IPPE), Obninsk 249020, Russia  
}

\date{Received: 2012-11-13 / Accepted: 2012-12-19}

\maketitle

\begin{abstract}
The radiolanthanide $^{161}$Ho (2.48 h) is a pro\-mising Auger-electron emitter for internal radiotherapy that can be produced with particle accelerators. The excitation functions of the $^{nat}$Dy(p,xn)$^{161}$Ho and $^{nat}$Dy(d,x)\-$^{161}$Ho reactions were measured up to 40 and 50 MeV respectively by using the stacked foil activation method and  -ray spectrometry. The experimental data were compared with results of the TALYS code available in the TENDL 2011 library \cite{1}. The main parameters of different production routes are discussed. 
\end{abstract}

\section{Introduction}
\label{intro}
The radiolanthanide $^{161}$Ho is an Auger-electron emitter having also low energy photons in high abundance. It is very suitable for internal radiotherapy of small tumors because of the low energy electrons emitted. The short range of Auger electrons however requires that labeled compounds approach the cell nucleus. It is also interesting as low-energy narrow-band X-ray source for internal irradiation \cite{2,3,4,5,6}.
Different routes exist to produce $^{161}$Ho with particle accelerators. One route is  $\alpha$- or $^3$He- particle irradiation of $^{159}$Tb relying on the $^{159}$Tb($\alpha$,2n)$^{161}$Ho and $^{159}$Tb($^3$He,n)$^{161}$Ho reactions. Another route is the irradiation of dysprosium targets using protons via the $^{161}$Dy(p,n), $^{162}$Dy(p,2n) or deuterons via $^{160}$Dy(d,n) and $^{161}$Dy(d,2n) reactions. When using the $^{159}$Tb($\alpha$,2n)\-$^{161}$Ho, $^{162}$Dy(p,2n) or $^{161}$Dy(d,2n) reactions processes, emission of one neutron also takes place resulting in simultaneous production of $^{162}$Ho, which is a radionuclide impurity. This radionuclide has a short half-life ground state $^{162g}$Ho (15 min, I$^{\pi}$  = 1$^+$) and a longer-lived isomeric state (67.0 min, I$^{\pi}$  = 6$^-$). From the point of view of $^{161}$Ho production, the contamination with the longer-lived excited state has some importance at the beginning after EOB, which will be reduced by the waiting time. The decay through internal transition of $^{162m}$Ho is followed only by low energy low intensity  $\gamma$-ray emission, but the 38\% electron capture decay results in strong high energy  $\gamma$-lines (see Table 1).
The cross-sections of the $^{159}$Tb($\alpha$,2n)$^{161}$Ho reaction was investigated by several authors (see in comparison of production routes). Although the basic cross-section data are still missing the proton induced reaction was also was used for practical production \cite{5}. The deuteron induced reaction was not used yet for the production and no cross-section data were published. We decided to investigate the excitation functions of the proton and deuteron routes experimentally. 
Naturally occurring dysprosium is composed of 7 stable isotopes ($^{156}$Dy - 0.06\%, $^{158}$Dy - 0.10\%, $^{160}$Dy - 2.34\%, $^{161}$Dy - 18.9\%, $^{162}$Dy - 25.5\%, $^{163}$Dy - 24.9\% and $^{164}$Dy - 28.2\%). Taking into account that many of these stable Dy isotopes have a larger mass number than $^{161}$Ho, the direct experimental investigation of the $^{161}$Dy(p,n), $^{161}$Dy(d,2n) or $^{160}$Dy(d,n) reactions, and routes leading to possibly disturbing activation products, should  require highly enriched targets.
A second possibility is to use natural targets and to use results of theoretical calculations to separate the contributions of the different target isotopes. This supposes an accurate predictivity of the calculations and a method to check the reliability has to be implemented.
We adopted this approach by making measurement of production cross-section of $^{161}$Ho on $^{nat}$Dy target with protons and deuterons and by comparing our experimental data with the predictions of the theoretical model codes. In case of good agreement we can then compare the different charged particle production routes using theoretical results validated by integral experiment.  

\section{Experimental and data evaluation}
\label{sec:2}
The general characteristics and procedures for irradiation, activity assessment and data evaluation (including estimation of uncertainties) were similar as in our many earlier works \cite{7,8,9,10,11}.
The main experimental parameters for the present study including the chosen monitor reactions \cite{12} are summarized in Table 1. The main methods used in data evaluation and the used decay data \cite{8,13,14,15,16,17,18,19,20} are collected in Table 2 and Table 3.
The excitation function of simultaneously measured proton and deuteron monitor reactions and comparison with recommended values are shown in Fig. 1.

\begin{table}
\tiny
\caption{\textbf{Main experimental parameters}}
\label{tab:1}       
\begin{tabular}{|p{0.9in}|p{0.9in}|p{0.9in}|} \hline 
\textbf{Reaction} & \textbf{${}^{nat}$Dy(p,x)} & \textbf{${}^{nat}$Dy(d,x)} \\ \hline 
 &  &  \\ \hline 
Incident particle & Proton  & Deuteron  \\ \hline 
Method  & Stacked foil & Stacked foil \\ \hline 
Target and thickness  & ${}^{nat}$Dy foil, 100.59 $\mu$m & ${}^{nat}$Dy foil, 100.59 $\mu$m \\ \hline 
Number of target foils & 15 & 15 \\ \hline 
Accelerator & CGR 560 cyclotron of Vrije Universiteit Brussels & Cyclone 90 cyclotron of the Université Catholique in CityplaceLouvain la Neuve (LLN)  \\ \hline 
Primary energy & 36 MeV & 50 MeV \\ \hline 
Irradiation time & 71 min & 30 min \\ \hline 
Beam current & 61 nA & 120 nA \\ \hline 
Monitor reaction, [recommended values]  & ${}^{nat}$Ti(p,x)${}^{48}$V reaction \cite{12} & ${}^{27}$Al(d,x)${}^{24}$Na  reaction \cite{12} \\ \hline 
Monitor target and thickness & ${}^{nat}$Ti, 10.9 $\mu$m & ${}^{nat}$Al, 26.96 $\mu$m \\ \hline 
detector & HpGe & HpGe \\ \hline 
$\gamma$-spectra measurements & 3 series & 3 series \\ \hline 
Cooling times & 1.5 h, 20 h, 80 h & 4h, 20h, 120h \\ \hline 
\end{tabular}
\end{table}

\begin{figure}
  \includegraphics[width=0.5\textwidth]{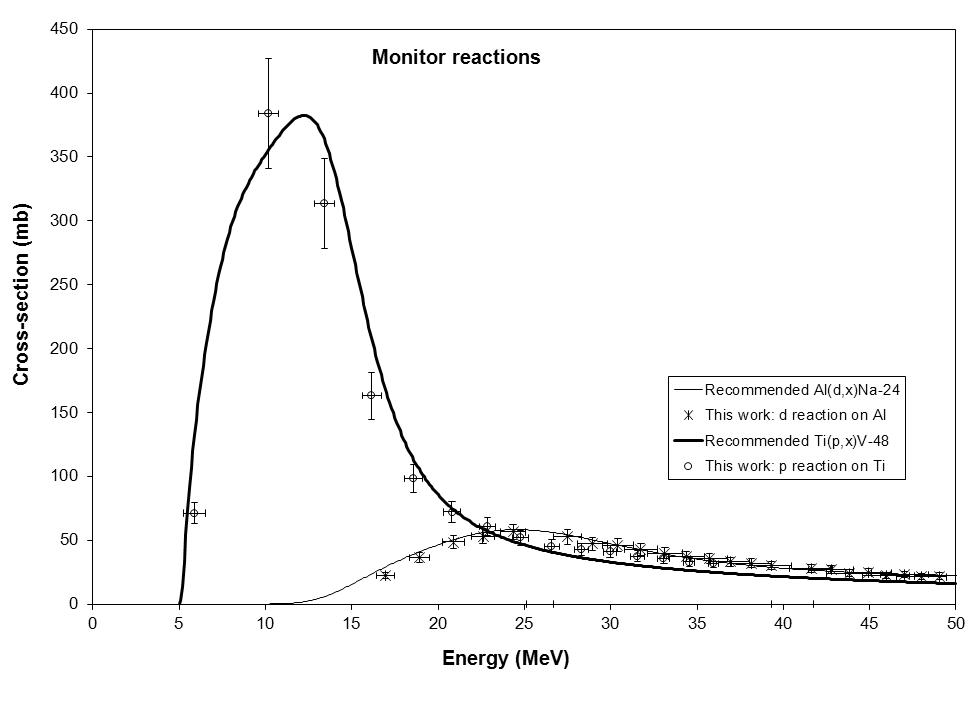}
\caption{The simultaneously measured monitor reactions for determination of proton beam energy and intensity}
\label{fig:1}       
\end{figure}

\begin{table}
\tiny
\caption{\textbf{Main parameters of data evaluation (with references)}}
\label{tab:2}       
\begin{tabular}{|p{1.1in}|p{1.1in}|p{0.4in}|} \hline 
$\gamma$-spectrum evaluation & Genie 2000, Forgamma & \cite{13,14} \\ \hline 
Determination of beam intensity & Faraday cup (preliminary)\newline Fitted monitor reaction (final) & \cite{15} \\ \hline 
Decay data & NUDAT 2.6 & \cite{16} \\ \hline 
Reaction Q-values & Q-value calculator & \cite{17} \\ \hline 
Determination of  beam energy & Andersen (preliminary)\newline Fitted monitor reaction (final) & \cite{18}\newline \cite{8} \\ \hline 
Uncertainty of energy & cumulative effects of possible uncertainties & \cite{19} \\ \hline 
Cross-sections & Isotopic cross section & [\cite{20} \\ \hline 
Uncertainty of cross-sections & Sum in quadrature of all individual contributions & \cite{19} \\ \hline 
Yield & Physical yield & \cite{20} \\ \hline 
\end{tabular}
\end{table}

\section{Results}
\label{sec:3}

\subsection{Cross-sections}
\label{sec:3.1}
The measured excitation functions for $^{nat}$Dy(p,xn)\-$^{161,162m}$\-Ho and $^{nat}$Dy(d,xn)$^{161,162m}$Ho are shown in Figs 2-3 and 5-6 in comparison with the results of the model calculations. The numerical data important for further data evaluation are collected in Tables 4 and 5. 
In both cases the theoretical results reproduce exceptionally well the shape of the measured excitation functions, but an overestimation over the whole energy range is seen for the $^{nat}$Dy(p,xn)$^{161,162m}$Ho and $^{nat}$Dy(d,xn)$^{161}$Ho reactions. In case of the $^{nat}$Dy(d,xn)$^{162m}$Ho the agreement of the maximum is acceptable.  The numerical values of theoretical results by a factor of 0.7 in case of  $^x$Dy(p,xn)\-$^{161,162m}$Ho and of 0.55 in case of $^x$Dy(d,xn)$^{161}$\-Ho should be multiplied as a rough estimation. There is no normalization for $^x$Dy(d,xn)$^{162m}$Ho reaction.
For further discussion we have normalized the theoretical cross-sections of the contributing reactions involved in $^{161}$Ho production with these factors. 
The comparison of the normalized  TENDL 2011 cross-sections of the $^{161}$Dy(p,n)\-$^{161}$Ho and the $^{162}$Dy\-(p,2n)\-$^{161}$Ho reactions and the $^{162}$\-Dy(p,n)$^{162m}$Ho  impurity reaction  are shown in Fig. 4 and of the $^{160}$Dy(d,n)\-$^{161}$Ho, the $^{161}$Dy(d,2n)$^{161}$\-Ho reactions and the $^{161}$Dy(d,n)$^{162m}$Ho  impurity reaction  is presented in  Fig 7.

\begin{table}
\tiny
\caption{\textbf{Decay characteristics of the ${}^{161}$Ho and ${}^{162}$Ho and Q-values of reactions for their productions)}}
\label{tab:3}       
\begin{tabular}{|p{0.45in}|p{0.3in}|p{0.3in}|p{0.3in}|p{0.5in}|p{0.45in}|} \hline 
Nuclide & Half-life & E${}_{\gamma}$(keV) & I${}_{\gamma}$(\%) & Contributing reaction & Q-value\newline (keV) \\ \hline 
\textbf{${}^{161}$Ho\newline }$\varepsilon $: 100 \%\newline 7/2-\textbf{} & 2.48 h & 77.42\newline 103.05\newline 157.26\newline 175.42 & 1.9\newline 103.05\newline 0.49\newline 0.43 & ${}^{161}$Dy(p,n)\newline ${}^{162}$Dy(p,2n)\newline ${}^{163}$Dy(p,3n)\newline ${}^{164}$Dy(p,4n) & -1640.64\newline -9837.63\newline -16108.65\newline -23766.77 \\ 
\textbf{} &  &  &  & ${}^{160}$Dy(d,n)\newline ${}^{161}$Dy(d,2n)\newline ${}^{162}$Dy(d,3n)\newline ${}^{163}$Dy(d,4n)\newline ${}^{164}$Dy(d,5n) & 2589.18\newline -3865.2\newline -12062.2\newline -18333.21\newline -25991.33 \\ \hline 
\textbf{${}^{162m}$Ho\newline }IT: 62 \%\newline $\varepsilon $: 38 \%\textbf{\newline }6-\newline 105.87 keV\textbf{\newline } & 67.0 m & 184.99\newline 282.86\newline 937.17\newline 1220.04 & 23.94\newline 10.27\newline 10.44\newline 23.7 & ${}^{162}$Dy(p,n)\newline ${}^{163}$Dy(p,2n)\newline ${}^{164}$Dy(p,3n)\newline  & -3027.91\newline -9298.92\newline -16957.05\newline  \\ 
\textbf{} &  &  &  & ${}^{161}$Dy(d,n)\newline ${}^{162}$Dy(d,2n)\newline ${}^{163}$Dy(d,3n)\newline ${}^{164}$Dy(d,4n) & 2944.51\newline -5252.48\newline -11523.49\newline -19181.61 \\ \hline 
\textbf{${}^{162g}$Ho\newline }$\varepsilon $: 100 \%\textbf{\newline }1+~\textbf{\newline } & 15.0 m & ~80.7\newline 1319.75 & 8.0\newline 3.82 & ${}^{162}$Dy(p,n)\newline ${}^{163}$Dy(p,2n)\newline ${}^{164}$Dy(p,3n)\newline  & -2922.04\newline ~~-9193.05\newline -16851.18 \\ 
\textbf{} &  &  &  & ${}^{161}$Dy(d,n)\newline ${}^{162}$Dy(d,2n)\newline ${}^{163}$Dy(d,3n)\newline ${}^{164}$Dy(d,4n) & ~3050.38\newline ~-5146.61\newline -11417.62\newline -19075.74 \\ \hline 
\end{tabular}
\end{table}

\begin{figure}
  \includegraphics[width=0.5\textwidth]{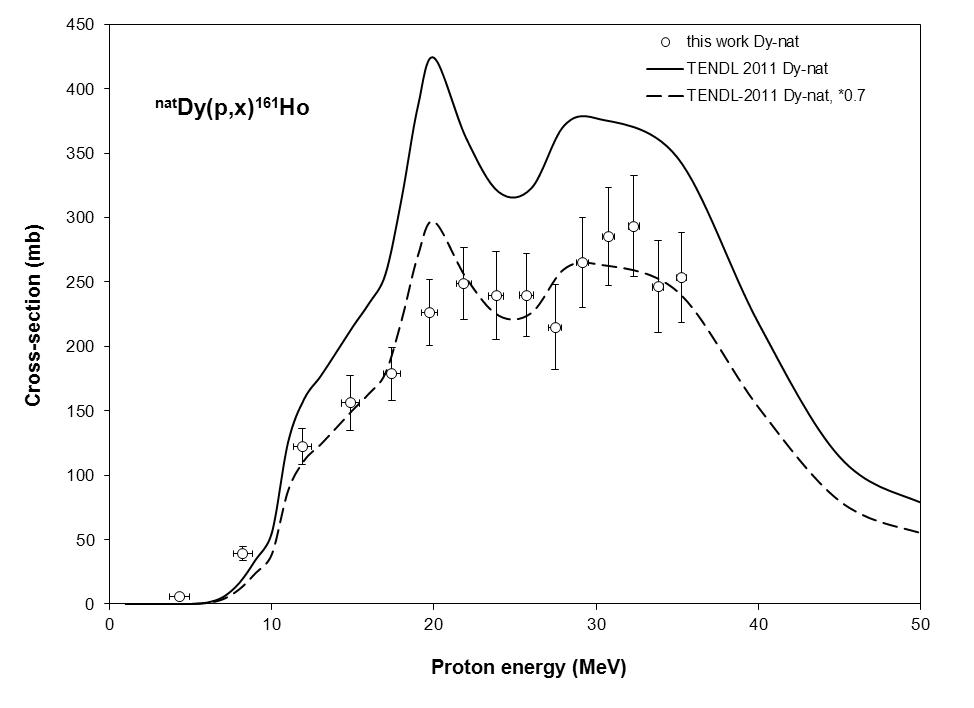}
\caption{Experimental cross-sections of the $^{nat}$Dy(p,xn)$^{161}$Ho  reaction in comparison with the results  of model calculations in TENDL 2011}
\label{fig:2}       
\end{figure}

\begin{figure}
  \includegraphics[width=0.5\textwidth]{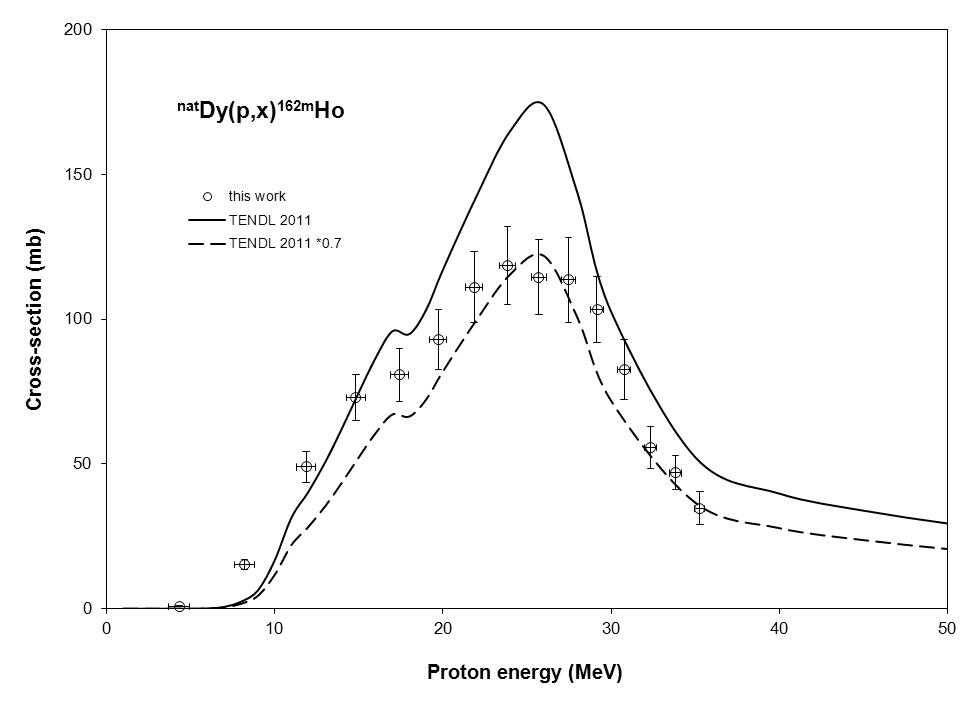}
\caption{Experimental cross-sections of the $^{nat}$Dy(p,xn)$^{162m}$Ho  reaction in comparison with the results  of model calculations in TENDL 2011}
\label{fig:3}       
\end{figure}

\begin{figure}
  \includegraphics[width=0.5\textwidth]{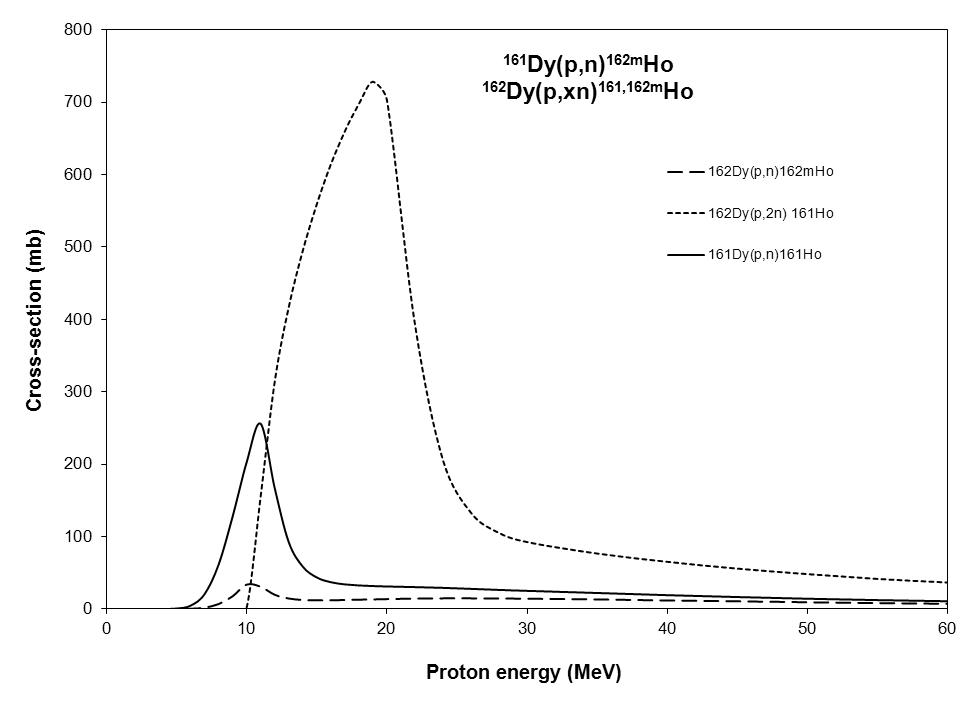}
\caption{Experimental cross-sections of the $^{nat}$Dy(d,xn)$^{161}$Ho reaction in comparison with the results of model calculations in TENDL 2011}
\label{fig:4}       
\end{figure}

\begin{figure}
  \includegraphics[width=0.5\textwidth]{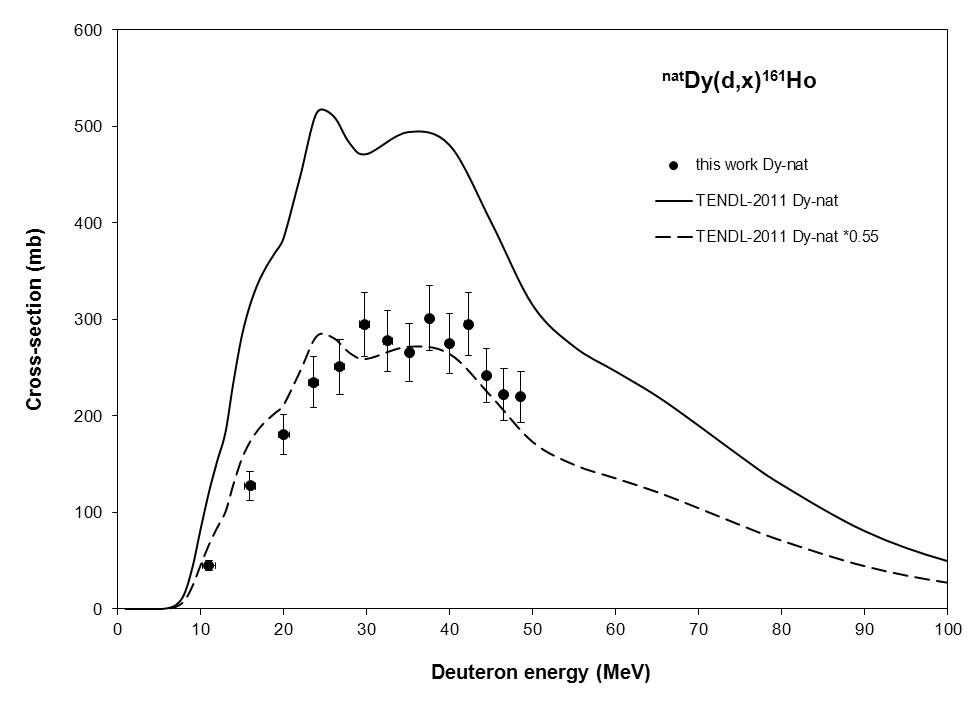}
\caption{Experimental cross-sections of the $^{nat}$Dy(d,xn)$^{162m}$Ho reaction in comparison with the results of model calculations in TENDL 2011}
\label{fig:5}       
\end{figure}

\begin{table}
\tiny
\caption{\textbf{Measured cross-sections of the ${}^{nat}$Dy(p,x)${}^{161,162m}$Ho reactions}}
\label{tab:4}       
\begin{tabular}{|p{0.4in}|p{0.4in}|p{0.4in}|p{0.4in}|p{0.4in}|} \hline 
 & \multicolumn{2}{|p{0.8in}|}{\textbf{${}^{1}$${}^{61}$Ho}} & \multicolumn{2}{|p{0.8in}|}{\textbf{${}^{1}$${}^{62m}$Ho${}^{ }$}} \\ \hline 
\textbf{E} &  $\sigma$ & $\pm\delta\sigma$ & $\sigma$ & $\pm\delta\sigma$ \\ \hline 
\textbf{MeV} & \multicolumn{2}{|p{0.8in}|}{\textbf{mb}} & \multicolumn{2}{|p{0.8in}|}{\textbf{mb}} \\ \hline 
35.3 &   253.5 & 34.8 & 34.7 & 5.7 \\ \hline 
33.8 &   246.2 & 35.6 & 46.9 & 5.9 \\ \hline 
32.3 &   293.4 & 39.0 & 55.6 & 7.3 \\ \hline 
30.8 &   285.5 & 38.1 & 82.6 & 10.2 \\ \hline 
29.2 &   265.0 & 35.2 & 103.4 & 11.5 \\ \hline 
27.5 &   214.8 & 33.2 & 113.6 & 14.6 \\ \hline 
25.7 &   239.8 & 32.2 & 114.6 & 13.1 \\ \hline 
23.8 &   239.2 & 34.1 & 118.6 & 13.4 \\ \hline 
21.9 &   248.9 & 28.1 & 111.1 & 12.3 \\ \hline 
19.7 &   226.1 & 25.5 & 93.0 & 10.4 \\ \hline 
17.4 &   178.8 & 20.6 & 80.8 & 9.0 \\ \hline 
14.8 &   156.1 & 21.1 & 73.0 & 8.0 \\ \hline 
11.9 &   122.3 & 14.1 & 49.0 & 5.4 \\ \hline 
8.2 &   39.2 & 5.8 & 15.3 & 1.7 \\ \hline 
4.3 &   5.7 & 1.7 & 0.8 & 0.1 \\ \hline 
\end{tabular}
\end{table}

\begin{table}
\tiny
\caption{\textbf{Measured cross-sections of the ${}^{nat}$Dy(d,x)${}^{161,162m}$Ho reactions}}
\label{tab:5}       
\begin{tabular}{|p{0.3in}|p{0.4in}|p{0.4in}|p{0.4in}|p{0.4in}|} \hline 
 &   \multicolumn{2}{|p{0.8in}|}{\textbf{${}^{1}$${}^{61}$Ho}} & \multicolumn{2}{|p{0.8in}|}{\textbf{${}^{1}$${}^{62m}$Ho${}^{ }$}} \\ \hline 
\textbf{E} &  $\sigma$ & $\pm\delta\sigma$ & $\sigma$ & $\pm\delta\sigma$ \\ \hline 
\textbf{MeV} & \multicolumn{2}{|p{0.8in}|}{\textbf{mb}} & \multicolumn{2}{|p{0.8in}|}{\textbf{mb}} \\ \hline 48.6 &   219.9 & 26.6 & 66.7 & 9.0 \\ \hline 
46.5 &   222.2 & 26.9 & 75.2 & 10.0 \\ \hline 
44.4 &   242.0 & 28.1 & 94.4 & 11.2 \\ \hline 
42.3 &   295.2 & 33.1 & 93.9 & 10.8 \\ \hline 
40.0 &   274.8 & 31.3 & 106.1 & 12.5 \\ \hline 
37.6 &   301.0 & 33.7 & 143.2 & 15.7 \\ \hline 
35.1 &   265.8 & 29.8 & 166.8 & 18.4 \\ \hline 
32.5 &   277.7 & 31.3 & 192.5 & 21.1 \\ \hline 
29.8 &   295.1 & 33.1 & 189.9 & 20.8 \\ \hline 
26.8 &   250.9 & 28.3 & 172.8 & 19.0 \\ \hline 
23.6 &   234.9 & 26.3 & 147.5 & 16.3 \\ \hline 
20.0 &   181.0 & 20.5 & 126.1 & 13.9 \\ \hline 
16.0 &   127.5 & 14.7 & 91.7 & 10.1 \\ \hline 
11.0 &   45.0 & 5.5 & 36.6 & 4.2 \\ \hline 
\end{tabular}
\end{table}

\subsection{Integral yields}
\label{sec:3.2}

The integral yields calculated on the basis of the normalized  TENDL 2011 cross-sections for the $^{nat}$Dy(p,xn)\-$^{161}$Ho,  $^{nat}$Dy(p,xn)$^{162m}$Ho, $^{nat}$Dy(d,xn)$^{162m}$Ho and $^{nat}$\-Dy\-(d,xn)$^{162m}$Ho production reactions are shown in Fig. 8. The calculated integral yield represents so called physical yield i.e. yield obtained in a short irradiation \cite{19}. The $^{nat}$Dy(p,xn)$^{161}$Ho yields are compared with the experimental data of Stephens \cite{5}. The value calculated by the Stephens' result for 11.6 MeV proton bombardment is significantly lower than our result, but it can be caused by the fact that the irradiation time was not published in that paper. Comparing the saturation activities, which are 1.8 GBq by Stephens (after 3.6 hours) and 1.89 GBq in our measurement/calculation, the agreement can be considered as good.

\begin{figure}
  \includegraphics[width=0.5\textwidth]{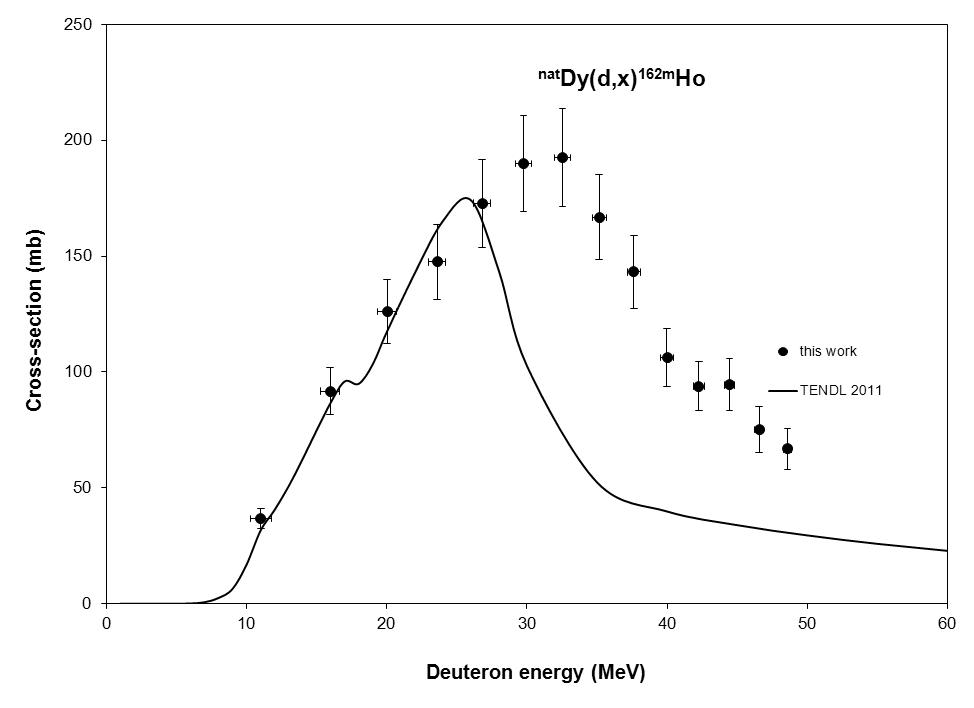}
\caption{Comparison of the cross-sections of the $^{161}$Dy(p,n)$^{161}$Ho and the $^{162}$Dy(p,2n)$^{161}$Ho reactions and the $^{162}$Dy(p,n)$^{162m}$Ho  impurity reaction in TENDL 2011}
\label{fig:6}       
\end{figure}

\begin{figure}
  \includegraphics[width=0.5\textwidth]{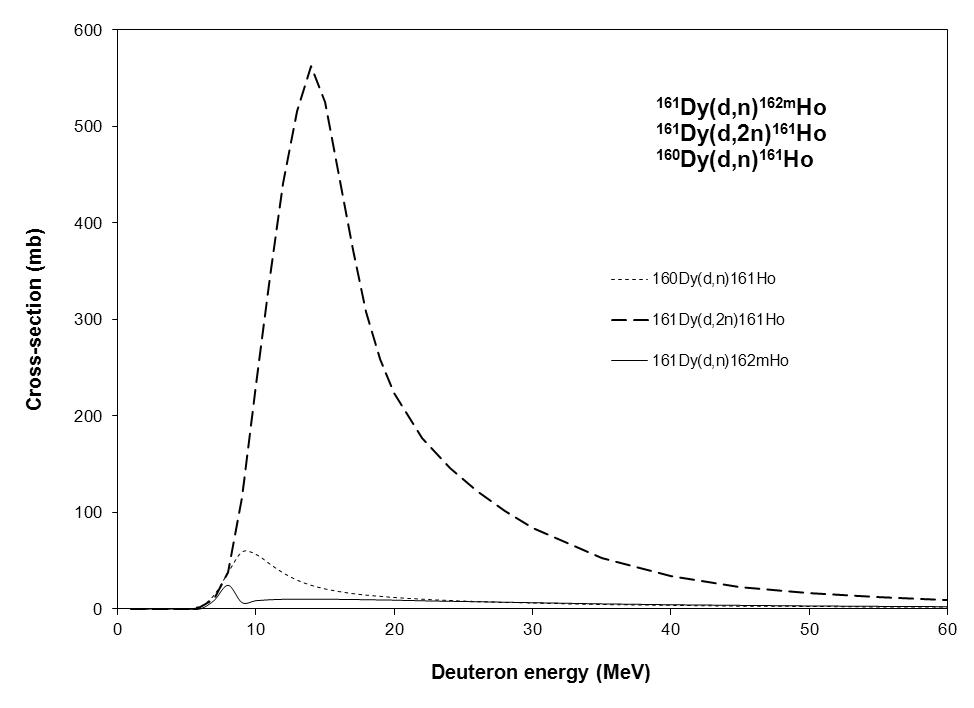}
\caption{Comparison of the cross-sections of the $^{160}$Dy(d,n)$^{161}$Ho and the $^{161}$Dy(d,2n)$^{161}$Ho reactions and the $^{161}$Dy(d,n)$^{162m}$Ho  impurity reaction in TENDL 2011}
\label{fig:7}       
\end{figure}

\begin{figure}
  \includegraphics[width=0.5\textwidth]{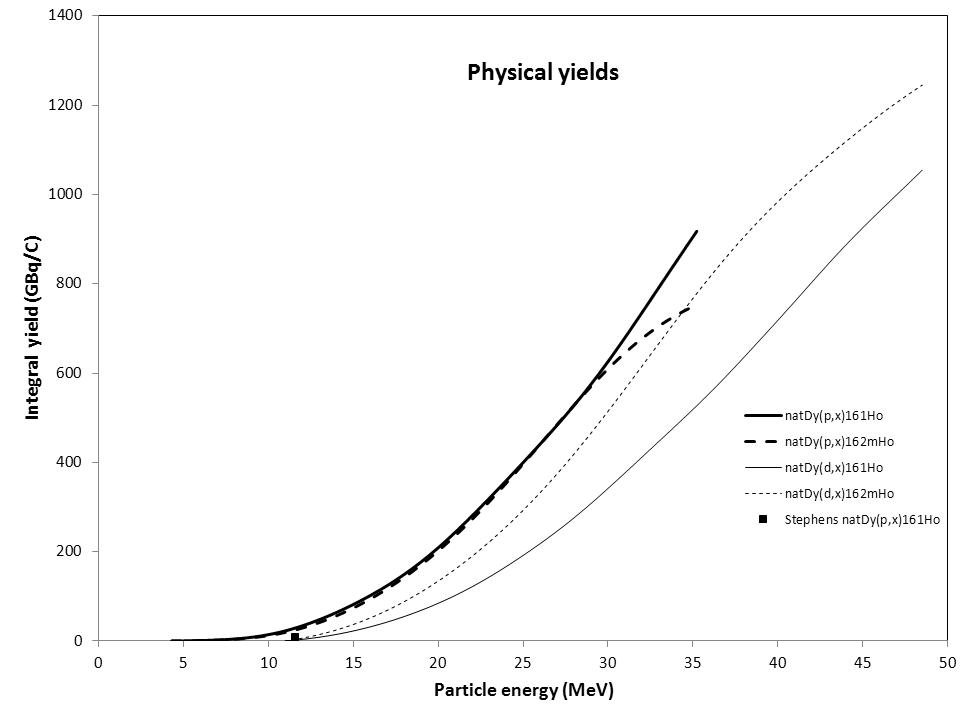}
\caption{Integral yields of the  $^{nat}$Dy(p,xn)$^{161}$Ho, $^{nat}$Dy(p,xn)$^{162m}$Ho, $^{nat}$Dy(d,xn)$^{162m}$Ho  and $^{nat}$Dy(d,xn)$^{162m}$Ho reactions}
\label{fig:8}       
\end{figure}

\section{Comparison of production routes on different target materials}
\label{sec:4}
The main parameters of the selected low and medium energy reactions that can lead to production of $^{161}$Ho on different target materials are collected in Table 6. The excitation functions of the proton and deuteron routes are shown in Figs. 2-3 and Fig 4-5. Mukherjee \cite{21} and Bonesso \cite{22} reported earlier experimental cross-section data on $^{159}$Tb($\alpha$,2n)$^{161}$Ho and Mukherjee \cite{21} and Singh \cite{23} on total cross-section of and $^{159}$Tb($\alpha$,2n)$^{162}$Ho. The $^{162m}$Ho/$^{162g}$Ho isomeric ratio was measured by Tulinov \cite{24}, Baskova \cite{25}. No experimental data were found for the $^{159}$Tb($^3$He,n)$^{161}$Ho reaction. The experimental data from literature and the theoretical excitation functions of the $^{159}$Tb($\alpha$,2n)$^{161}$Ho and $^{159}$Tb($\alpha$,n)$^{162m}$Ho reactions are shown in Fig. 9 and 10 respectively. The  theoretical excitation functions of the $^{159}$Tb($\alpha$,2n)$^{161}$Ho and $^{159}$Tb(,n)$^{162m}$Ho  and the $^{159}$Tb($^3$He,n)$^{161}$Ho reactions are compared in Fig. 11. From the excitation functions of the above mentioned reactions the following conclusions can be drawn:

\begin{itemize}
\item	The production yields for the $^{162}$Dy(p,2n) is the highest  followed by the $^{161}$Dy(d,2n), $^{159}$Tb($\alpha$,2n) and $^{161}$Dy(p,n)
\item	No $^{162m}$Ho impurity is produced when using of $^{159}$Tb\-($^3$He,n), $^{161}$Dy(p,n) and the $^{160}$Dy(d,n) reactions.  Among them the $^{161}$Dy(p,n)  reaction has the  highest cross-section ( max~260 mb) followed by the  $^{160}$Dy(d,n)  reaction( max~60 mb) and the less productive  $^{159}$Tb\-($^3$He,n) ( max~1 mb) 
\item	The element Tb is monoisotopic, relatively cheap and recovery is practically not necessary
\item	In case of proton and deuteron induced reactions highly enriched 160, 161 or 162 Dy targets are required 
\item	The production cross-sections of the $^{162m}$Ho from $^{169}$Tb($\alpha$,n), $^{161}$Dy(d,n) and $^{162}$Dy(d,n) are low
\item	The impurity level depends on the selected energy range. The ratio of the saturation yields of the main reaction and of competing impurity reaction is shown in Fig. 12 as a function of energy. In the production energy range the ratio is lower than 3%
\item	The half-life of $^{162m}$Ho is three times shorter, therefore by using a short irradiation, the activity impurity level will reach 3\%. But by using irradiations lasting two half-life of $^{161}$Ho and taking into account 1 hour needed for the chemical separation and the labeling process the impurity level of $^{162m}$Ho will be reduced to 1\% by decay.
\end{itemize} 

\begin{figure}
  \includegraphics[width=0.5\textwidth]{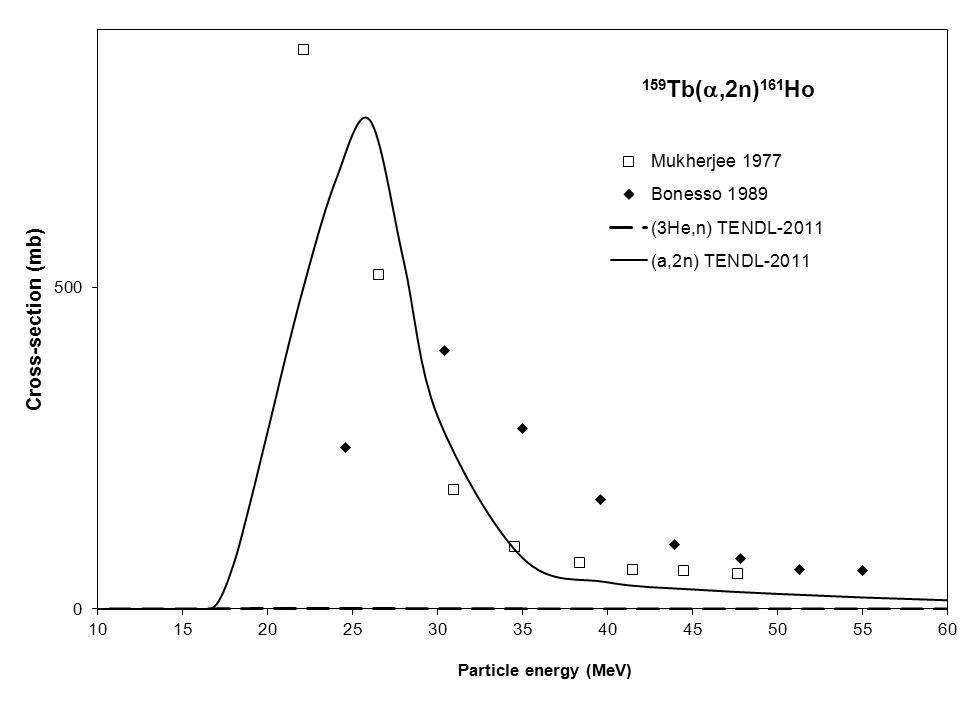}
\caption{Experimental and   theoretical cross-sections of the $^{159}$Tb($\alpha$,2n)$^{161}$Ho reaction}
\label{fig:9}       
\end{figure}

\begin{figure}
  \includegraphics[width=0.5\textwidth]{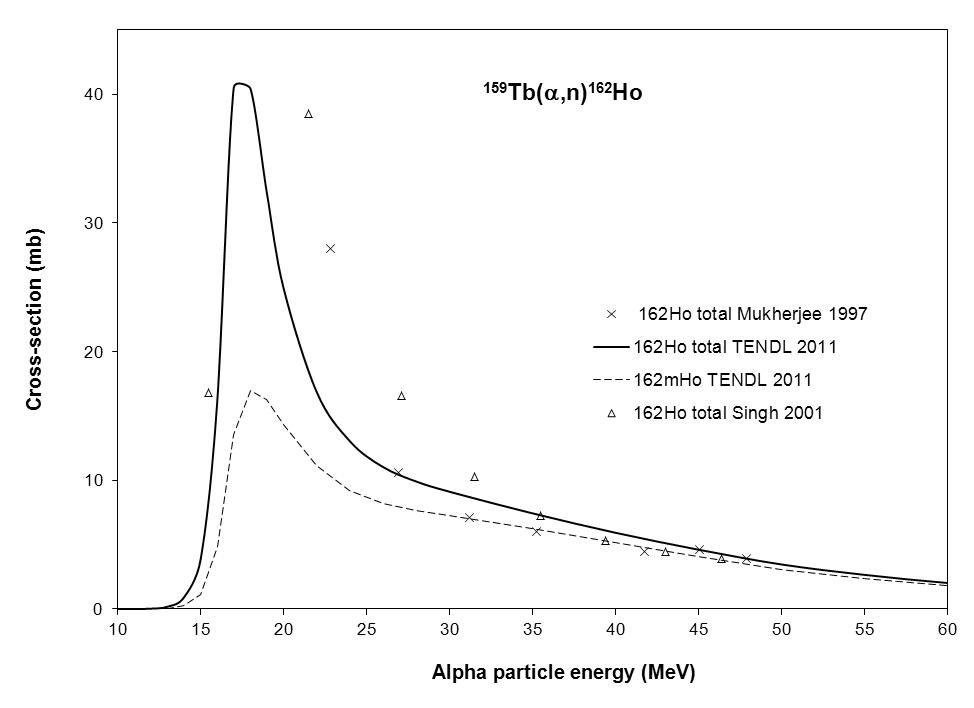}
\caption{Experimental and   theoretical cross-sections of the $^{159}$Tb($\alpha$,2n)$^{162m}$Ho reaction}
\label{fig:10}       
\end{figure}
 
\section{Summary and conclusions}
\label{sec:5}
The principal aim of this work was an investigation of the production possibility of the radiotherapy related $^{161}$Ho. We present first experimental cross-sections for  $^{nat}$Dy(p,xn)$^{161}$Ho, $^{nat}$Dy(p,xn)$^{162m}$Ho, $^{nat}$Dy(d,x)$^{161}$Ho and $^{nat}$Dy(d,x)$^{162m}$Ho up to 40 and 50 MeV incident particle energies respectively. 
The TENDL 2011 theoretical data predict well the shape of the excitation functions but overestimate the absolute values with a nearly constant factor in the whole energy range. The comparison of the different production routes shows that for production of $^{161}$Ho with high radionuclide purity the $^{161}$Dy(p,n)$^{161}$Ho, $^{162}$Dy (p,2n)  and $^{161}$Dy(d,2n) reactions  give the highest production yields.  The $^{162m}$Ho radionuclide impurity level of the last two reactions however is significant. No enriched target material is necessary in case of $^{159}$Tb($\alpha$,2n) (Tb is monoisotopic) but it requires accelerators having medium energy alpha particles. The $^{159}$Tb($^3$He,n) and $^{160}$Dy(d,n) reactions have very low cross-sections, and accelerators disposing of $^3$He beam are rare and the $^3$He irradiation without recovery of $^3$He gas is expensive. 
On the basis of the production yields, the impurity levels and the requirements of the medical application the $^{161}$Dy(p,n)  reaction is the  production method of best choice.

\begin{table*}
\tiny
\caption{\textbf{Summary of the production parameters for selected reactions}}
\label{tab:6}       
\begin{tabular}{|p{0.9in}|p{0.5in}|p{0.8in}|p{0.6in}|p{0.5in}|p{0.45in}|p{0.65in}|p{0.5in}|p{0.4in}|} \hline 
\textbf{Reaction} & \textbf{Q-value} & \textbf{Impurity reaction} & \textbf{Optimal\newline  energy range} \newline \textbf{(MeV)} & \textbf{${}^{161}$Ho\newline thick target yield\newline (GBq/C)} & \textbf{Impurity\newline level\newline (\%)} & \textbf{Optimal\newline  energy range\newline at low impurity\newline (MeV)} & \textbf{${}^{161}$Ho\newline thick target yield\newline (GBq/C)} & \textbf{Impurity\newline level\newline (\%)} \\ \hline 
${}^{159}$Tb($\alpha$,2n)${}^{161}$Ho & -16053.93 & ${}^{159}$Tb($\alpha$,n)${}^{162}$${}^{m}$Ho & 35-19 & 165 & 6 & 30-23 & 107 & 2.8 \\ \hline 
${}^{159}$Tb(${}^{3}$He,n)${}^{161}$Ho   & ~4523.7 & no & 30-15 &  & 0 &  &  &  \\ \hline 
${}^{161}$Dy(p,n)${}^{161}$Ho & -1640.64 & no & 15-8 & 132 & 0 &  &  &  \\ \hline 
${}^{162}$Dy(p,2n)${}^{ 161}$Ho & -9837.63 & ${}^{162}$Dy(p,n)${}^{ 162}$${}^{m}$Ho & 30-12 & 1459 & 8.5 & 22-15 & 868 & 4.6 \\ \hline 
${}^{160}$Dy(d,n)${}^{161}$Ho & 2589.18 & no & 15-7 & 34 & 0 &  &  &  \\ \hline 
${}^{161}$Dy(d,2n)${}^{161}$Ho & -3865.2 & ${}^{161}$Dy(d,n)${}^{ 162}$${}^{m}$Ho & 30-16 & 1454 & 1.8 & 28-20 & 955 & 1.5 \\ \hline 
\end{tabular}
\end{table*}

\begin{figure}
  \includegraphics[width=0.5\textwidth]{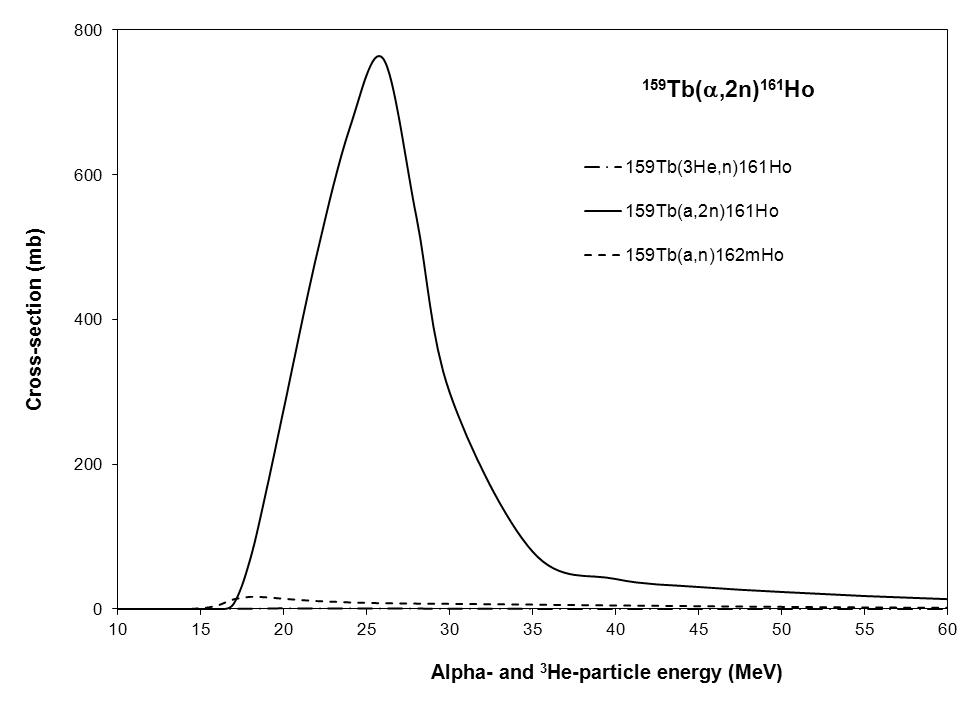}
\caption{Comparison of the cross-sections of the $^{159}$Tb($\alpha$,2n)$^{161}$Ho and the $^{159}$Tb($^3$He,n)$^{161}$Ho reactions and the $^{159}$Tb($\alpha$,n)$^{162m}$Ho impurity reaction (TENDL 2011)}
\label{fig:11}       
\end{figure}

\begin{figure}
  \includegraphics[width=0.5\textwidth]{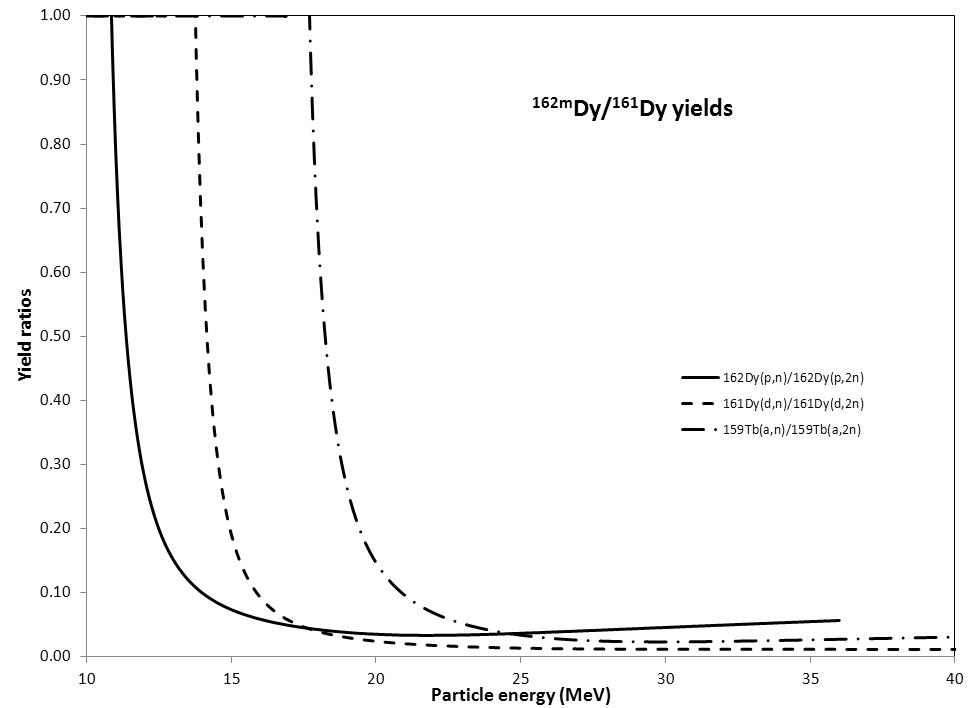}
\caption{12.	The ratio of the saturation yield of the main reaction and the satellite impurity reaction}
\label{fig:12}       
\end{figure}

\begin{acknowledgements}
This work was performed in the frame of the HAS-FWO Vlaanderen (Hungary-Belgium) project. The authors acknowledge the support of the research project and of the respective institutions in providing the beam time and experimental facilities.
\end{acknowledgements}

\bibliographystyle{spphys}       
\bibliography{Ho161}   

%
%

\end{document}